# Partition Quantitative Assessment (PQA): A quantitative methodology to assess the embedded noise in clustered omics and systems biology data

**Camacho-Hernández, Diego A.**[1,2]†, **Nieto-Caballero, Victor E.**[1,2]†, **León-Burguete, José E.**[1,2], and **Freyre-González, Julio A.**[1,*]

[1] Regulatory Systems Biology Research Group, Laboratory of Systems and Synthetic Biology and
[2] Undergraduate Program in Genomic Sciences, Center for Genomic Sciences, Universidad Nacional Autónoma de México (UNAM), Morelos, Mexico.

† These authors contributed equally to this work.
* Corresponding author: jfreyre@ccg.unam.mx

**Abstract:** Identifying groups that share common features among datasets through clustering analysis is a typical problem in many fields of science, particularly in post-omics and systems biology research. In respect of this, quantifying how a measure can cluster or organize intrinsic groups is important since currently there is no statistical evaluation of how ordered is, or how much noise is embedded in the resulting clustered vector. Many of the literature focuses on how well the clustering algorithm orders the data, with several measures regarding external and internal statistical measures; but none measure has been developed to statistically quantify the noise in an arranged vector posterior a clustering algorithm, i.e., how much of the clustering is due to randomness. Here, we present a quantitative methodology, based on autocorrelation, to assess this problem.

## 1. Introduction

A common task in today's research is the identification of specific markers, as predictors of a classification yielded in clustering analysis of the data. For instance, this approach is particularly useful after high-throughput experiments to compare gene expression or methylation profiles among different cell lines [1]. This task is coming handful in the nascent field of single-cell sequencing, leading the important step of clustering cells to further classification or as a qualifying metric of the sequencing process [2]. Regarding the vastly used gene expression assays, the vector of profiles for each marker across different cell lines is recorded using hierarchical clustering algorithms. These algorithms yield a dendrogram and a heat map representing the vector of marker profiles, illustrating the arrangement of the clusters. To assess how well the clustering is segregating different cell lines, a class stating the desired partitioning of each cell line is provided *a posteriori*. Then, a simple visual inspection of the vector of classes is used to estimate whether the clustering is providing a good partition. Such partition vector is colored according to the classification that each item is associated with, and it is expected that similar items will be contiguous, so the formed groups are assessed qualitatively on the biological background of each item.

This procedure should not be confused with "supervised clustering", which provides a vector of classes starting the desired partitioning *a priori*. This is then used to guide the clustering algorithms by allowing the learning of the metric distances that optimizes the partitioning [3]. Additionally, it may get confused with the metric assessment of the clustering algorithms, especially with the external cluster evaluation. For this, various metrics have been developed to qualify the clustering algorithm itself, such as intrinsic and extrinsic measures. These metrics are used for clustering algorithm validation. The extrinsic validation compares the clustering to a goal to say whether it is a good clustering or not. The internal validation compares the elements within the cluster and their differences [4]. PQA involves characteristics of both kinds of validation, through using both the crafted goal standard and the yielded signal itself (clustered vector). However, PQA gathers these elements not qualifying the clustering algorithm itself but to quantify the noise embedded in the cluster, this noise may be due to the intrinsic metric or marker used to order the data set.

A possible caveat of the qualitative assessment discussed above is that humans tend to perceive meaningful patterns within random data leading to a cognitive bias known as apophenia [5]. While interpreting the partitions obtained from unsupervised clustering analysis, researchers attempt to visually assess how close the classifications are to each other finding patterns that are not well supported by the data. Such an effect is raised because the adjacency between items may give a notion of the dissimilarity distance in the dendrogram leaves. Unfortunately, as much as we know, there is no method to quantitatively assess the quality of the groups of classifications from the clustering or, at least, there is no attempt to quantify whether certain configuration or order of the items may be due to randomness. This is a serious caveat, since the insertion of noise can lead to false conclusion or misleading results. Furthermore, the purging of this noise can lead to a more efficient descriptions of markers and its phenomena, accelerating the advance in many fields.

In statistics, serial correlation (SC) is a term used to describe the relationship between observations of the same variable over specific periods. It was originally used in engineering to determine how a signal, for instance, a radio wave, varies with itself over time. Later, SC was adapted to econometrics to analyze economic data over time principally to predict stock prices and, in other fields, to model-independent random variables [6]. We applied the SC to propose a manner to quantify how well is the grouping of a posterior classification just by retrieving the results of unsupervised clustering analysis. Thus, we propose a novel relative score, PQA, to solve the subjectivity of the visual inspection and to statistically quantify how much noise is embedded in the results of clustering analysis.

## 2. Methodology

2.1. Assigning numeric labels to classifications

A vector denoting the putative similarities among the variables in a study is usually obtained after a clustering analysis. Each variable is classified to generate a vector of profiles (VP). Such a vector of classifications is usually translated into a colors vector, in which each color represents a classification. It is common to inspect this vector to find groups that make sense according to the analyzed data. To the method presented in this work, the VP may be as simple as a vector of strings or numbers that represent the input.

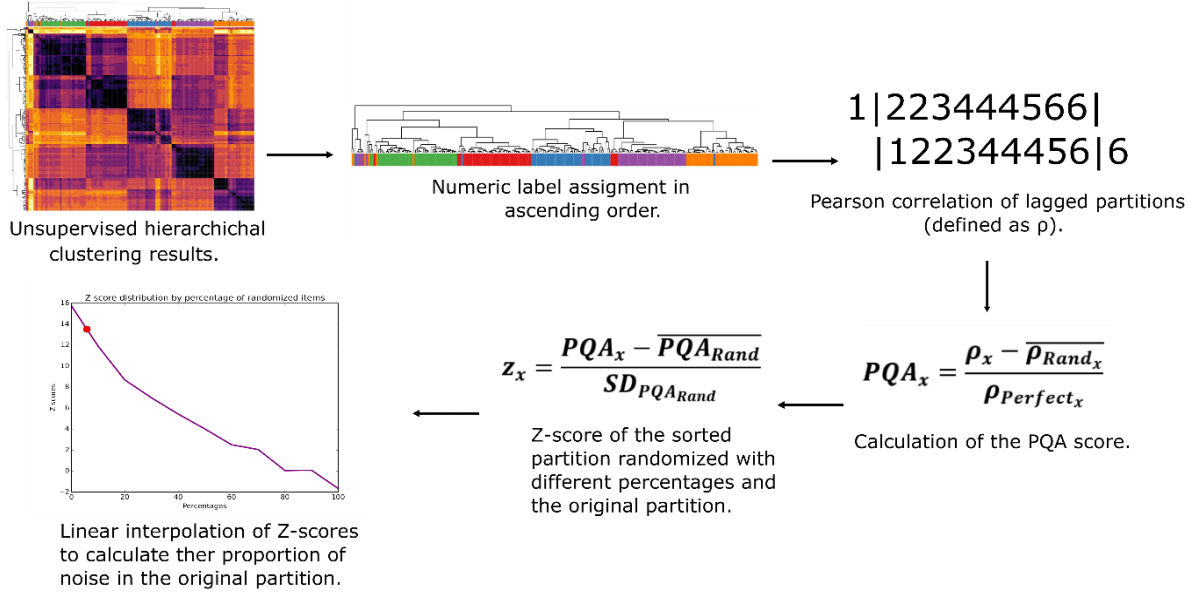

**Figure 1.** The pipeline of the PQA methodology.

Whatever representation of the classifications may be, it is necessary to transform the classifications to a vector of numeric labels, in which a number represents a classification, to be able to calculate SC. To accomplish this, we assign the first numeric label (number 1) to the first item in the vector, which usually lays at one of the vector's extremes. Then, if the classification o the next item is different from the previous one, the next number in the sequence is assigned, and so on. This way of labeling warrants that the changes in the SC values are due to the order of numbers, that is to say, the grouping of the classifications resulting from the clustering, and it is not an artifact of the labeling itself (Figure 1).

2.2. PQA score

Because the order of the VP could be interpreted as the grouping of the classifications, we measure how well the same classifications are held together in the VP through an SC shifted one position. Such sort of correlation is defined as the Pearson-product-moment correlation between the VP discarding the first item, and the VP discarding the last (Equation 1, $x_i$ (order vector i-th position), n (length of x), $\rho_i$ (resulting SC)).

$$\rho_i = \frac{\sum_{i=2}^{n}\left(x_i - \frac{\sum_{j=2}^{n} x_i}{n-1}\right) \sum_{i=1}^{n-1}\left(x_i - \frac{\sum_{j=1}^{n-1} x_i}{n-1}\right)}{\sqrt{\sum_{i=2}^{n}\left(x_i - \frac{\sum_{j=2}^{n} x_i}{n-1}\right)^2} \sqrt{\sum_{i=1}^{n-1}\left(x_i - \frac{\sum_{j=1}^{n-1} x_i}{n-1}\right)^2}} \quad (1)$$

We then define the PQA as the SC of the VP after removing background noise, normalized for the SC of the percent grouping partitions (defined as the sorted vector in ascending order). This, the more similar VP is to its sorted vector, the higher the score is yielded (Equation 2, $\boldsymbol{\rho_x}$ (SC of the VP), $\overline{\rho_{Rand_x}}$ (Mean of the SC of one thousand randomizations), $\boldsymbol{\rho_{Perfect_x}}$ SC of the sorted vector in ascending order)).

$$PQA_x = \frac{\rho_x - \overline{\rho_{Rand_x}}}{\rho_{Perfect_x}} \quad (2)$$

2.3. Background-noise correlation factor in the PQA score

To compute the background-noise correlation factor in the PQA score definition, we sample the indexes of the VP and the swapping the corresponding items. This background correction is aimed to remove inherent noise in the data, even though the score may still be subjected to noise from the chosen clustering algorithm or discrepancies in the posterior classification.

2.4. Statistical significance of the PQA score

To quantify the statistical significance of the PQA score, we calculate a Z-score (Equation 3),

$$Z_x = \frac{PQA_x - \overline{PQA_{Rand}}}{SD_{PQA_{Rand}}} \quad (3)$$

where $PQA_x$ is the PQA score of the VP, $\overline{PQA_{Rand}}$ is the mean of PQA scores of one thousand randomizations of the VP. These randomizations have the purpose of generating a solid random background to compare it to the real signal. The number of randomizations does not depend on the size of the VP. It is worth to notice that there are two randomization processes, one is meant to generate the input population of random vectors to calculate the PQA score to further calculate a Z-score and the other is representing the noise in Equation 2.

2.5. Defining noise proportions

To provide a quantification of the embedded noise in the VP, we calculate the Z-scores from the distribution of PQA values of the randomized vectors. This shuffling is yielded by scrambling the vector. Then this Z-score is interpolated to retrieve the estimated noise in the VP cluster.

2.6. Effect of the length and number of partitions of the vector in the Z-score distributions.

Since we want to compare the PQA with the noise, we randomized 1000 times the VP. We opted to describe the dynamic of the Z-score given the different percentage of noise and the number of partitions. For this, we synthetically crafted vector of both ranging from 0 to 100 elements and number of classifications. The Z-scores were retrieved from the crafted vectors using the formulas described above.

**3. Results and Discussion**

3.1. Effects of permuted numeric labels on the partition

We wondered whether the correct assigning of numeric labels to alter the less possible the SC calculations, so we analyzed how the SC changes over the synthetic partitions with permuted labels. We began generating synthetic partitions in ascending and descending order, increasing both the number of classifications and the number of items, up to 100. It is important to highlight that the number of items belonging to each classification was kept constant. Because trying all the possible permutations for each vector would be implausible, we created a subset of 1000 permutations of each vector, then we calculated the mean SC (Figure 1, see Methodology). We observed that the mean SC got high when the number of items in the VP was greater or equal to 2 times the number of classifications, nevertheless, we got the highest SC when the numeric labels we assigned by sequential order, either ascending or descending (Figure 2).

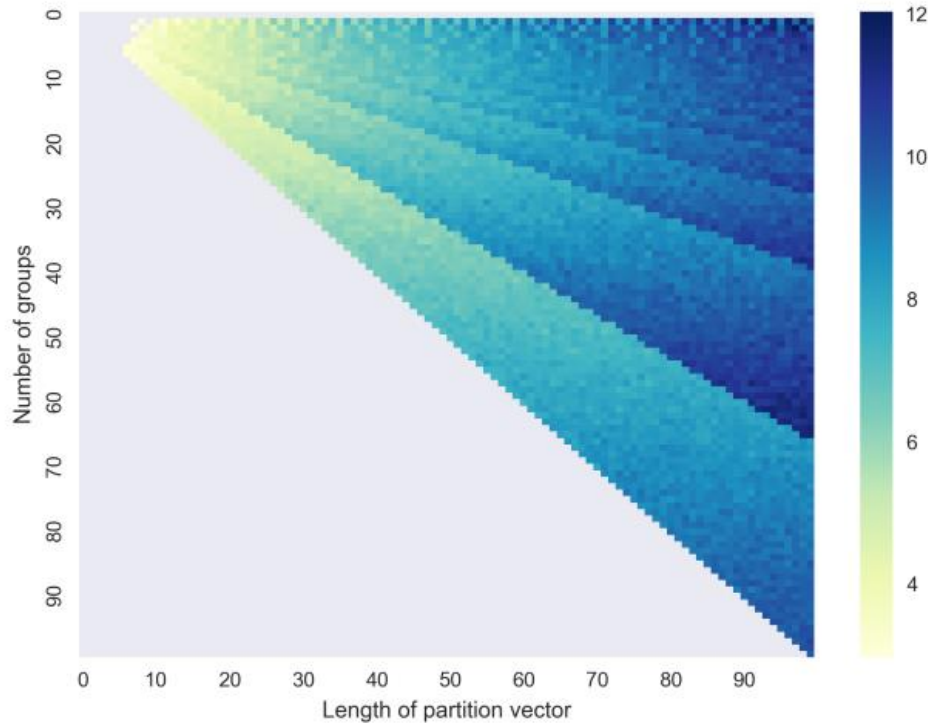

**Figure 2.** Z-scores of the PQA scores from partitions varying in the number of classifications and the length of the partition.

3.2. Length of partitions as a proxy of the number of classifications

We wonder whether the number of classifications and the length of the VP may change the statistical significance of the PQA score because of the less the number of items in the VP, the greater the chance to group each item with any order. We then tested such effect by calculating a Z-score from ordered synthetic partitions increasing both the number of classifications and the number of items up to 100. We also kept constant the number of classifications for the sake of this analysis. We noticed that only the length of the partition has a true effect on the Z-score, but that is not the case for the number of classifications. We observed that every partition minor than 13 could be considered as pure noise, however, we consider a Z-score cutoff of greater than 3 (p-value of 0.002). We also observed Z-score values still greater than 2 with a length of 12, 11, and 10, but lesser than with lengths between 2 and 9 (Figure 2). If we were more flexible, we could have laid out a length cutoff on those values without losing statistical significance, since a Z-score of 2 corresponds roughly to a p-value of 0.05. The results of this analysis were expected by intuition because the probability of an item to occupy a position in the VP increases the number of items does the same.

3.3. Proof of concept: Quantifying real noise

After a literature revision, we noticed that some datasets were subject to visual inspection in their respective papers, so we applied our method to quantify the proportion of noise embedded in those datasets and to test whether they may lead to apophenia. We choose two datasets from literature because of two main reasons, first, the data should have a high number of items that are way above our Z-score significance threshold (>13) and, second, we wanted contrasting orderings of the partitions so to have one dataset that looks very disordered and another that looks somewhat ordered to compare the noise proportions. Lastly, we assessed the behavior of the metric in highly ordered data. This also matches our threshold mentioned above.

3.3.1. Cancer methylation signatures

The first dataset consists of methylation profiles of 242 different cancerous and non-cancerous samples [7] (Figure 3). Though the classifications look very sparse and the groups are torn apart in many subgroups distributed along with the data's VP. We detected 25.1% of noise and a PQA score of 0.53 (Figure 4, with a Z-score of 8.2 and a p-value of 9.6x10$^{-17}$), both numbers imply that even though there may be disordered in the VP, there is not a very high noise proportion nor a high PQA score. These results suggest that, like any other statistical test, the longer the number of items in the partition the more diluted is the effect of disorder in the VP, and the results also lead to a greater statistical significance as shown in the analysis of the number of items and classifications. Besides the authors concluded that their clustering analysis results made sense from their molecular and biological background, as well as the perspectives about the analyzed profiles, they only assessed grouping just by visual inspection and concluded the grouping was well done. However, understanding the noise in the cluster can help to pursue better markers since it could help to narrow the search space in these kinds of studies.

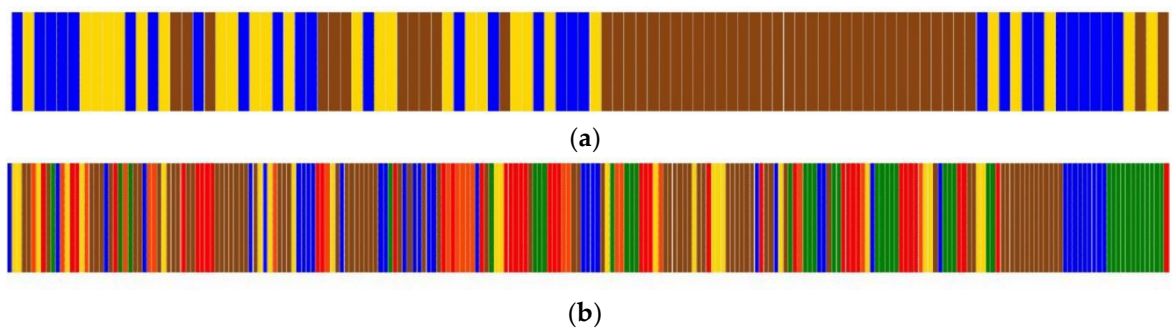

**Figure 3.** Visual representation of clustered data used to assess the method. (**a**) Dataset from Jie Shen et. al. (**b**) Dataset from Tooyoka et. al.

3.3.2. Distribution of microRNAs in cancer

The second dataset consists of 103 expression profiles of microRNAs from three classes of samples: invasive breast cancer, those with ductal carcinoma in situ (DCIS), and health (Figure 3) [8]. The authors visually identified three clusters, though selecting the right cutting height threshold is difficult. Besides, one of the clusters is a mix of classes in different proportions, leading the authors to arguably conclude that the DCIS and control sample profiles are not different. On this matter, the PQA score and the proportion of noise are 0.62 and 30.2%, respectively (Figure 4, with Z-score of 6.2 and a p-value of 3.9x10$^{-10}$) providing a quantitative assay to support the grouping that the authors claimed. Furthermore, in comparison with the methylation profiles discussed above, we can appreciate that a partition which appear even less fuzzy has even a higher noise ratio, supporting the idea of how visual inspection could lead to misleading results.

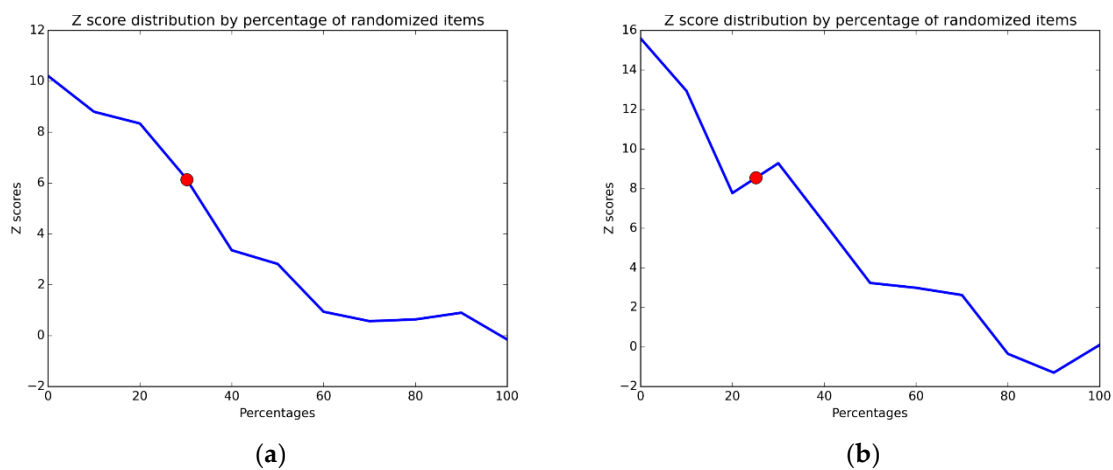

**Figure 4.** Z-score distribution by percentage of randomized items. (**a**) Dataset from Jie Shen et. al. (**b**) Dataset from Tooyoka et. al. The red dots represent the Z-score interpolation of the corresponding data sets.

3.3.3. Comparison of genetic regulatory networks with theoretical models

Finally, to assess the PQA methodology using systems biology data we clustered 210 networks according to their pairwise dissimilarity [9]. First, 42 curated biological networks were retrieved from Abasy Atlas (v2.2) [10]. For each biological network, we then constructed four networks each according to a theoretical model (Barabasi-Alberts, Erdos-Renyi, Scale-free, and Hierarchical-modular). We estimated the parameters of each theoretical model from the properties of the corresponding biological network. The models used reproduce one or more intrinsic characteristics of the biological networks, such as power-law distribution, hubs, and scale-free degrees, and hierarchical modular structure [11]. Visual inspection suggested that the classification yielded a highly ordered PV, distinguishing according to the nature of each network (Figure 5). The PQA score for this VP is 0.92 (p-value = $2.5 \times 10^{-40}$, Z-score =13.2) and the proportion of noise was 5.8% (Figure 6). In contrast to the previous examples, here we obtained a highly ordered clustering and a very low proportion of noise, which suggests that although the models recapitulate some of the properties of genetic regulatory networks, each of them is not sufficient to capture their structural properties.

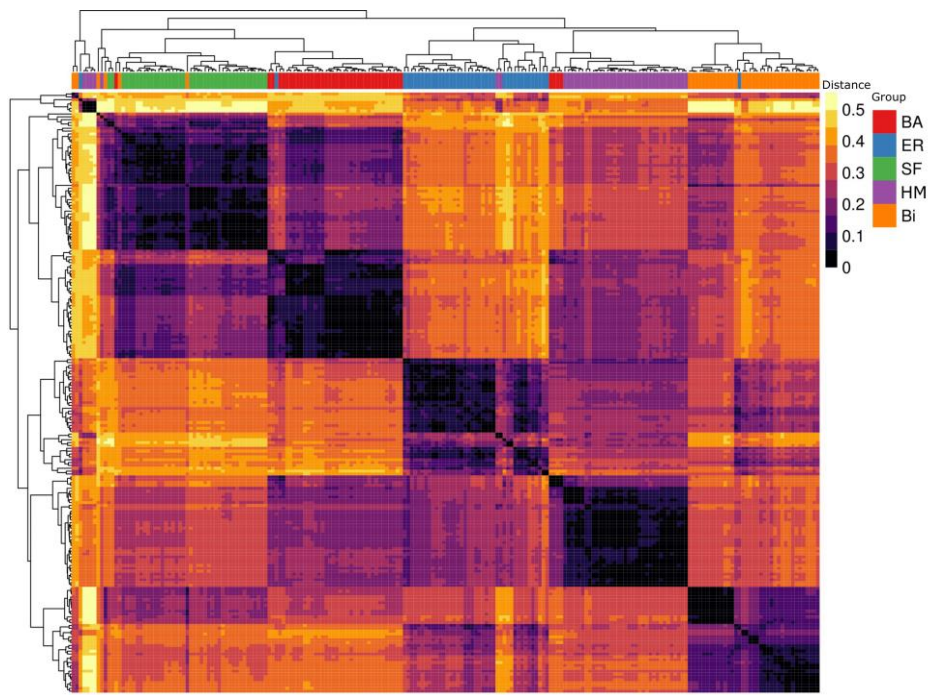

**Figure 5.** Cluster analysis of distance among gene regulatory networks and theoretical network models. The abbreviations and colors used in the posterior classification are as follows: Barabasi-Alberts (BA, red), Erdos-Renyi (ER, blue), Scale-free (SF, green), Hierarchical modularity (HM, purple), and biological networks (Bi, orange).

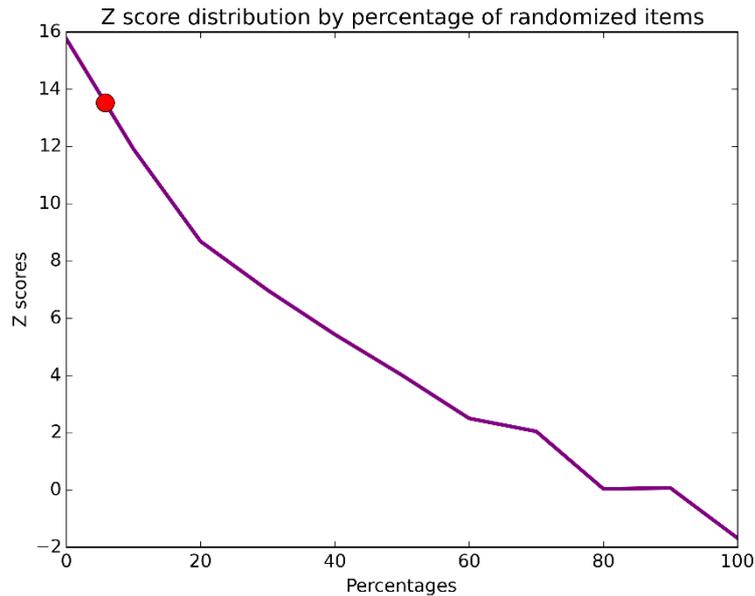

**Figure 6.** Z-score distribution by percentage of randomized items of VP from genetic regulatory networks. The red dot represents the Z-score interpolation of the actual data set.

## 4. Conclusions

In this work, we presented a novel method to quantify the proportion of noise embedded in the grouping of associated classes of the elements in hierarchical clustering. We proposed a relative score derived from an SC of the VP from the dendrogram of any clustering analysis and calculated Z-statistics as well as an extrapolation to deliver an estimation of noise in the VP. We explain how the method is formulated and show the tests we made to systematically refine it.

We additionally made a proof of concept by using clustering data from two works that we think perfectly represent overfitting by apophenia. Additionally, we added an example from network biology where clustered networks are separated by intrinsic characteristics. Although in this work we focused on examples where hierarchical clustering is performed, this framework can apply to any partition algorithm in which the elements are identified and a vector of the order can be acquired.

We concluded that the clustered sets of biologic data have a high measure of noise, despite looking well grouped. We proved what a minimum number of classifications should be considered in this sort of clustering analysis to have a significant reduction of noise. On the other hand, we permuted the labels of the associated classes and concluded that the effect is negligible. We proved that randomness still plays an important role by biasing the results, though it may not be evident through visual inspection.

The PQA could be used as a benchmark to test what clustering algorithm should be appropriate for the analyzed dataset by minimizing the noise proportion and to guide omics experimental designs. Nevertheless, a word of caution, the PQA score alone can be subject to subjectivity if not used properly since it depended on the characteristics of the analyzed data. Thus, the PQA score is thought to be considered as a quantification of noise in clustered data and should be used with discretion.

**Author Contributions:** Conceptualization, J.A.F.G.; methodology, J.A.F.G.; software, D.A.C.H., V.E.N.C., and J.A.F.G.; validation, D.A.C.H., V.E.N.C., and J.A.F.G.; formal analysis, D.A.C.H., V.E.N.C., and J.A.F.G.; investigation, D.A.C.H., V.E.N.C., J.R.L.B., and J.A.F.G.; resources, J.A.F.G.; data curation, D.A.C.H., V.E.N.C., and J.E.L.B.; writing—original draft preparation, D.A.C.H., V.E.N.C., J.E.L.B., and J.A.F.G.; writing—review and editing, D.A.C.H., V.E.N.C., and J.A.F.G.; visualization, D.A.C.H., V.E.N.C., J.E.L.B., and J.A.F.G.; supervision, J.A.F.G.; project administration, J.A.F.G.; funding acquisition, J.A.F.G. All authors have read and agreed to the published version of the manuscript.

**Funding:** This work was supported by the Programa de Apoyo a Proyectos de Investigación e Innovación Tecnológica (PAPIIT-UNAM) [IN205918 to J.A.F.G.].

**Conflicts of Interest:** The authors declare no conflict of interest.